\documentclass{revtex4}
\usepackage{epsf}

\begin{document}
\title{Largely deformed states of $^{13}$B}

\author{Yoshiko Kanada-En'yo}
\address{Yukawa Institute for Theoretical Physics, Kyoto University,
Kyoto, Japan}

\author{Yusuke Kawanami\footnote{The present affiliation is TOHOKUSHINSHA FILM CORPORATION.},
 Yasutaka Taniguchi}
\address{Department of Physics, Kyoto University, Kyoto , Japan}
\author{Masaaki Kimura}
\address{Institute of Physics, University of Tsukuba, Tsukuba , Japan}

\begin{abstract}
The excited states of $^{13}$B were studied with a method of 
antisymmetrized molecular dynamics(AMD). 
The theoretical results suggest that 
the intruder states with large deformations
construct the rotational bands, $K^\pi=3/2^-$ and $K^\pi=1/2^+$, starting 
from 5 MeV and 8 MeV, respectively. The neutron structure of the
$K^\pi=3/2^-$ is analogous to the intruder ground state of $^{12}$Be.
In the predicted $K^\pi=1/2^+$ band, we found very exotic structure 
with a proton intruder configuration.
This proton intruder state
has a larger deformation than superdeformation.
The band-head $1/2^+$
state is assigned to the $1/2^+$(4.83 MeV), 
which was experimentally suggested to be  
the proton intruder state because of the 
strong production via the proton-transfer to the $^{12}$Be($0^+$) state
in the $^4$He($^{12}$Be,$^{13}$B$\gamma$)$X$ experiments.

\end{abstract}
\maketitle

\noindent
%PACS numbers: 21.60.-n, 02.70.Ns, 21.10.Ky, 27.20.+n.

\section{Introduction}	

In the recent progress of experimental and theoretical 
researches of unstable nuclei, various exotic phenomena have been
discovered. As well known, one of the attractive subjects in neutron-rich 
nuclei is the breaking of neutron magic number such as $N=8$ and $N=20$
suggested in neutron-rich $p$-shell and $sd$-shell nuclei.
The breaking of the $N=8$ shell in $^{11}$Be has been well known for a
long time because of the parity inversion of the ground state, 
$J^\pi=1/2^+$. For $^{12}$Be, various experimental researches have been
recently achieved \cite{iwasaki00,navin,pain} to study the properties of 
the ground band, and the breaking of $N=8$ shell closure has been 
established. Concerning the mechanism of the $N=8$ shell breaking, 
the recent observation\cite{pain} of the 
significant $d$-wave component in the $^{12}$Be ground state 
is the direct probe for the deformation, which 
should be one of the essential factors
for the breaking in Be isotopes
suggested in theoretical calculations
\cite{otsuka96,ITAGAKI,Enyo-be11,Enyo-be12,Oertzen-rev}.
In the theoretical side, many kinds of 
microscopic calculations have been performed to investigate
neutron-rich nuclei with the breaking shell closure.
In case of neutron-rich Be isotopes, various properties have been 
successfully described by many groups from a point of view of 
cluster (see references in Ref.\cite{Oertzen-rev}).
As a result, it is considered that molecular orbital 
structure with large deformation is a key 
for the breaking of $N=8$ magic number in neutron-rich Be.

In the molecular orbital picture, 
Be isotopes are described by 2 $\alpha$ clusters 
and valence neutrons which occupy
the molecular orbitals formed around the 
$2\alpha$ core\cite{ITAGAKI,Oertzen-rev,Okabe77,SEYA,OERTZEN}. 
The molecular orbitals are given by a linear combination of $p$-orbitals around
the $\alpha$ clusters, and they are associated with the orbitals 
in two-center shell model\cite{twocenter}. In the spherical limit,
the negative-parity $\pi$-orbitals are the lowest for the valence neutrons.
On the other hand, with the development of the $2\alpha$ clustering,
so-called $\sigma$ orbital, which is the longitudinal positive-parity orbital,
comes down because its kinetic energy decreases. Finally, the inversion of the
$\pi$ and $\sigma$ orbitals occurs in a system with 
well-developed $2\alpha$ clustering. 
The inversion of the molecular orbitals corresponds to 
the parity inversion of the '$p$' and '$sd$' orbitals 
because the $\pi$ and $\sigma$ orbitals in the molecular orbital model 
are associated with the $p$ and $sd$ orbitals in the deformed shell model,
respectively. 
The intruder state of $^{12}$Be
is written by the configuration with two 
neutrons in the $\sigma$-orbital. If the $2\alpha$ structure develops 
enough in $^{12}$Be, the intruder state
may become the ground state.
This necessarily causes the significant mixture of 
$d$-wave component, which is consistent with the 
recent measurements \cite{pain}.
Moreover, most of the low-lying states in neutron-rich Be 
isotopes can be well described by the molecular orbital 
structure\cite{ITAGAKI,Enyo-be11,Enyo-be12,Oertzen-rev,SEYA,OERTZEN,ARAI,DOTE,Enyo-be10,OGAWA,ITO04},
and a variety of cluster states has been predicted in excited states
of $^{11}$Be and $^{12}$Be 
\cite{Enyo-be11,Enyo-be12,Oertzen-rev,OERTZEN,Ito00,Descouvemont01}.
These facts indicate that cluster aspect is one of essential 
features in light unstable nuclei as well as in light stable nuclei.
In particular, cluster structure is favored in neutron-rich Be, where 
a variety of exotic structure arises 
due to the the formation of 2 $\alpha$ clusters and the molecular orbitals.

Let us turn to structure of B isotopes, which have a proton number 
$Z=5$ larger by one than Be isotopes.
In contrast to the situation of $^{12}$Be, 
$^{13}$B is considered to have 
the normal ground state with a neutron $p$-shell closed configuration. 
Comparing with the intruder configuration of the ground state in $^{12}$Be,
this shows that the $N=8$ magic number is restored in $^{13}$B
due to the additional proton. Thus, the additional proton gives
drastic structure change of the ground state, however,
it is still natural to expect that intruder states
may appear in the excited states of $^{13}$B. 
If a cluster structure can develop also
in B as well as Be, the intruder states may lie 
in low excitation energy region, because such states 
can be stabilized in a similar way to the 
ground state of $^{12}$Be.
The intruder states may have large deformation and 
construct rotational bands.
Cluster features in neutron-rich B isotopes have been systematically
studied in a chain of B isotopes with a molecular orbital 
model(MO) \cite{SEYA} and a method of 
antisymmetrized molecular dynamics(AMD)\cite{ENYObc,ENYOsup}.
These studies have been concentrated on the structure of ground states,
and have shown that $^{13}$B is most spherical among B isotopes, 
while developed cluster structure has been predicted in  
further neutron-rich B like $^{15}$B and $^{17}$B. 
These results suggest 
a trend of two-center cluster structure 
in $Z=5$ systems.

In this paper, we investigated deformed states of $^{13}$B 
by performing microscopic calculations of 
the ground and excited states of $^{13}$B. 
We focused on their cluster aspect. In particular, 
the molecular orbital structure is our major interest.
In the present study we applied an AMD method, 
which is known to be a powerful approach to investigate
cluster structure of unstable nuclei\cite{ENYObc,ENYOsup,AMDrev}. 
The present method is the same as that applied to the studies of 
$^{10}$Be,$^{11}$Be, and $^{12}$Be\cite{Enyo-be11,Enyo-be12,Enyo-be10}. 
Namely, we 
performed variation after spin-parity projection 
within the framework of AMD\cite{Enyo-c12}. This method
has been proved to successfully describe various properties of the 
ground and excited states of Be isotopes.

The paper is organized as follows. In the next section, we briefly explain 
the theoretical method of the present work. Results and discussions are 
given in \ref{sec:results}. Finally, we give a summary in 
\ref{sec:summary}.

\section{Formulation} \label{sec:formulation}
We performed energy variation after 
spin parity projection(VAP) within the AMD model space, as was done
in the previous studies\cite{Enyo-be12,Enyo-be10,Enyo-c12}. 
The detailed formulation of the AMD method 
for nuclear structure study is described in 
\cite{ENYObc,ENYOsup,AMDrev,Enyo-c12}.
In particular, the formulation of the present calculations is basically 
the same as that described in \cite{Enyo-be12,Enyo-be10}.

An AMD wave function is a Slater determinant of Gaussian wave packets;
\begin{equation}
 \Phi_{\rm AMD}({\bf Z}) = \frac{1}{\sqrt{A!}} {\cal{A}} \{
  \varphi_1,\varphi_2,...,\varphi_A \},
\end{equation}
where the $i$th single-particle wave function is written by a product of
spatial($\phi$), intrinsic spin($\chi$) and isospin($\tau$) 
wave functions as,
\begin{eqnarray}
 \varphi_i&=& \phi_{{\bf X}_i}\chi_i\tau_i,\\
 \phi_{{\bf X}_i}({\bf r}_j) &\propto& 
\exp\bigl\{-\nu({\bf r}_j-\frac{{\bf X}_i}{\sqrt{\nu}})^2\bigr\},
\label{eq:spatial}\\
 \chi_i &=& (\frac{1}{2}+\xi_i)\chi_{\uparrow}
 + (\frac{1}{2}-\xi_i)\chi_{\downarrow}.
\end{eqnarray}
$\phi_{{\bf X}_i}$ and $\chi_i$ are spatial and spin functions, and 
$\tau_i$ is iso-spin
function which is fixed to be up(proton) or down(neutron). 
We used a width parameter $\nu=0.18$ fm$^{-2}$, which 
is chosen to be the optimum value for $^{13}$B.
Accordingly, an AMD wave function
is expressed by a set of variational parameters, ${\bf Z}\equiv 
\{{\bf X}_1,{\bf X}_2,\cdots, {\bf X}_A,\xi_1,\xi_2,\cdots,\xi_A \}$.

For the lowest $J^\pi$ state,
we varied the parameters ${\bf X}_i$ and $\xi_{i}$($i=1\sim A$) to
minimize the energy expectation value of the Hamiltonian,
$\langle \Phi|H|\Phi\rangle/\langle \Phi|\Phi\rangle$,
for the spin-parity projected AMD wave function;
$\Phi=P^{J\pi}_{MK'}\Phi_{\rm AMD}({\bf Z})$.
Here, $P^{J\pi}_{MK'}$ is the spin-parity projection operator.
Then we obtained the optimum solution of the parameter set;
${\bf Z}^{J\pi}_1$ for the lowest $J^\pi$ state.
The solution ${\bf Z}^{J\pi}_n$ 
for the $n$th $J^\pi$ state are calculated by varying ${\bf Z}$ 
so as to minimize the energy of 
the orthogonal component to the lower states.

After the VAP calculations for the $J^\pi_n$ states with respect to various
$J$, $n$ and $\pi=\pm$,
we obtained the optimum intrinsic wave functions,
$\Phi_{\rm AMD}({\bf Z}^{J\pi}_n)$, 
which approximately describe the corresponding $J^\pi_n$ states. 
After the VAP, we superposed the 
spin-parity eigen wave functions 
projected from all the obtained AMD wave functions. 
Namely, we determined
the final wave functions for the $J^\pi_n$ states as, 
\begin{equation}\label{eq:diago}
|J^\pi_n\rangle=\sum_{i,K} c^{J^\pi_n}(K,J_i,\pi_i,k_i) 
|P^{J\pi}_{MK}\Phi_{\rm AMD}({\bf Z}^{J_i\pi_i}_{k_i})\rangle.
\end{equation}

\section{Results} \label{sec:results}

We adopted the same effective nuclear interaction as that used
in the study of Be isotopes \cite{Enyo-be11,Enyo-be12}, 
which consists of the central force, the
spin-orbit force and the Coulomb force.
The Majorana parameters in the MV1 force
are $m=0.65$, and the strengths of the spin-orbit force
are $u_{I}=-u_{II}=3700$ MeV. The VAP calculations of AMD 
using these interactions reproduce well the breaking of neutron magic
number $N=8$ 
in $^{11}$Be and $^{12}$Be. With this interaction the calculated 
binding energies of $^{12}$Be and $^{13}$B are
61.9 MeV and 76.4 MeV, which underestimate the experimental
values, 68.6 MeV and 84.5 MeV, respectively, however, we adopted
this parametrization because the energy levels of the excited states 
in $^{10}$Be, $^{11}$Be and $^{12}$Be are reasonably   
reproduced.

The basis AMD wave functions were obtained by the VAP for the ground and
excited states of $^{13}$B. The number of the basis AMD wave functions
in the present calculations are 23. The initial wave function 
in the energy variation was randomly chosen for $J\le 5/2$ states. 
For $J \ge 7/2$ states, we started the variational calculation 
from the initial wave function projected from the obtained 
wave function $\Phi_{\rm AMD}({\bf Z}^{J'_i\pi_i}_{k_i})$ for the
$J'\le 5/2$ states.
These independent AMD wave functions were
superposed to calculate the final wave functions.

\subsection{Energies and deformation}
$^{13}$B is a nucleus with neutron magic number $N=8$, and its ground
state is the $3/2^-$ with normal configuration
of $p$-shell closure. 
Above the $3/2^-$ ground state of $^{13}$B, it is experimentally 
known that many states exist in the excitation 
energy $E_x\ge 3.5$ MeV region with high level density. 
Unfortunately, spins and parities of most of these states are 
unknown. Recently, the state at 4.83 MeV has been assigned to be a 
$1/2^+$ state by $^4$He($^{12}$Be,$^{13}$B$\gamma$)$X$ 
experiments\cite{Ota07}. Because of its strong production 
via proton-transfer to the $^{12}$Be($0^+$) state, this excited 
state is suggested to be a proton intruder state.

\begin{figure}[th]
\epsfxsize=8. cm
\centerline{\epsffile{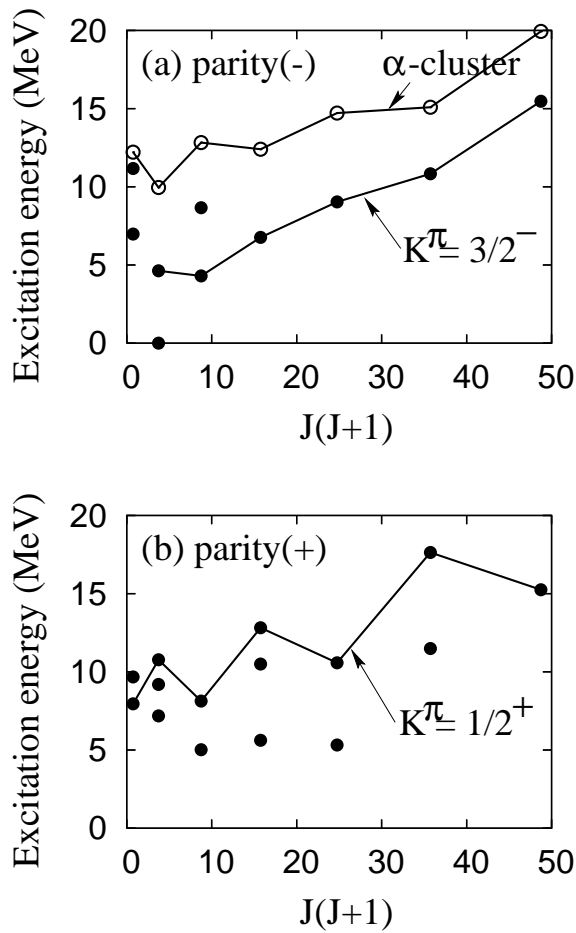}}
\vspace*{8pt}
\caption{Excitation energies of the negative- and positive-parity states 
of $^{13}$B calculated by superposition of the basis wave functions.
Filled circles are the energies of the 
$J^\pi_n$ states, for which the VAP calculations
were done. 
Open circles are the energies of the $J^\pi_n$ states, which were 
obtained by diagonalization of Hamiltonian by superposing 
wave functions, 
but the VAP calculations were not performed for the corresponding 
$J^\pi_n$ states. 
\protect\label{fig:b13spe}
}
\end{figure}

The calculated energy levels of the negative- and 
positive-parity states of $^{13}$B are shown in Fig.~\ref{fig:b13spe}. 
In addition to the ground $3/2^-_1$ state,
we obtained many excited states with various $J^\pi$
in the region $E_x\ge 4$ MeV.
These states may correspond to
the observed levels in this energy region.
In the excited states, we found 
three largely deformed bands, 
$K^\pi=3/2^-$, $K^\pi=1/2^+$ and $K^\pi=1/2^-$ (solid lines).
These bands are composed of intruder states 
or well-developed cluster states.
In particular, the band-head state $1/2^+_1$ of the $K^\pi=1/2^+$ is the
proton intruder state with a large deformation, and hence this 
should be assigned to the experimental $1/2^+$(4.83 MeV) state. 
The $K^\pi=1/2^-$ band was obtained by the spin-parity projection and the
diagonalization of the obtained wave functions, 
though the VAP calculations were not performed for the corresponding 
$J^\pi_n$ states. This band is dominantly the $\alpha$-$^9$Li cluster state. 
The intrinsic structures of these deformed states
are discussed later in detail.
As for other excited states (disconnected filled circles),
intrinsic deformation of 
the major AMD wave function $\Phi_{\rm AMD}({\bf Z}^{J\pi}_n)$, which dominates
the final wave functions $|J^\pi_n\rangle$, 
is small or as large as normal deformation at most. These excited states
are regarded to be dominated by $0\hbar\omega$ or 
and neutron $1\hbar\omega$ configurations.

\begin{figure}[th]
\epsfxsize=6 cm
\centerline{\epsffile{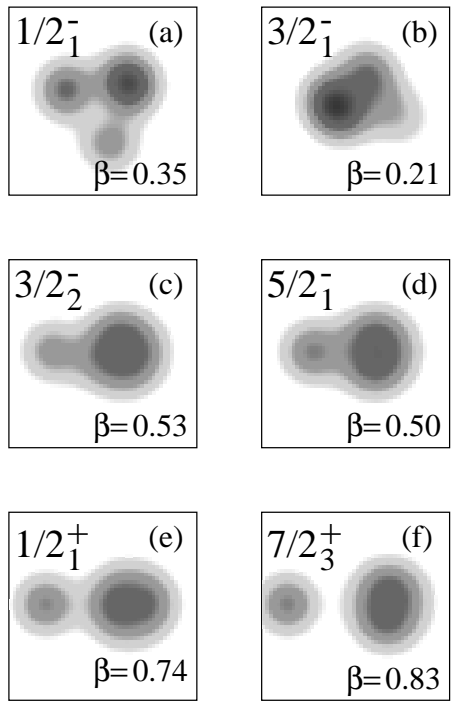}}
\vspace*{8pt}
\caption{
Density distribution of the ground and excited states of
$^{13}$B. 
The intrinsic density of the dominant AMD wave function 
in the $|J^\pi_n\rangle$ is shown. The density is integrated along the
axis perpendicular to the plane. 
The deformation parameter $\beta$\protect\cite{Enyo-oppo} of 
matter density for the intrinsic 
state are also given. The box size is 10 fm.
\label{fig:b13dense}
}
\end{figure}

Figure~\ref{fig:b13dense} shows density distribution and deformation parameters 
of the major AMD wave functions, $\Phi_{\rm AMD}({\bf Z}^{1/2-}_1)$,
$\Phi_{\rm AMD}({\bf Z}^{3/2-}_1)$, $\Phi_{\rm AMD}({\bf Z}^{3/2-}_2)$, 
$\Phi_{\rm AMD}({\bf Z}^{5/2-}_1)$, $\Phi_{\rm AMD}({\bf Z}^{1/2+}_1)$ and
$\Phi_{\rm AMD}({\bf Z}^{7/2+}_3)$, which were obtained by the VAP 
for the corresponding $J^\pi_n$ states. 
The ground state ($3/2^-_1$)
has the most spherical shape(Fig.~\ref{fig:b13dense}(b)),
 due to the neutron $p$-shell closure.
This is consistent with the previous work by AMD \cite{ENYObc}.
In the $1/2^-_1$ state(Fig.~\ref{fig:b13dense}(a)), 
a three-center cluster core structure appears.
The core clusters are an $\alpha$ with two valence neutrons, a triton
and an $\alpha$. This state is approximately regarded as the 
$SU(3)$-limit cluster state 
though the spatial cluster development 
is somehow contained. The similar three-center cluster structure 
is found also in the $5/2^-_2$ state.
In the $3/2^-_2$ and the $5/2^-_1$ states
(Fig.~\ref{fig:b13dense}(c) and (d)), 
we found remarkably deformed structure with 
developed cluster cores. These states are the members of the 
$K^\pi=3/2^-$ band which starts from $Ex=5$ MeV. 
It is interesting 
that such a largely deformed band appears only 5 MeV 
above the ground state,
 even though this nucleus has neutron magic number
$N=8$.
The $J^\pi=5/2^+_1$ state of this band is on the yrast line.
Moreover, it is striking that a further large deformation arises
in the $1/2^+_1$ state(Fig.~\ref{fig:b13dense}(e)), which is the band-head state 
of the $K^\pi=1/2^+$ band. The deformation $\beta=0.74$ 
of this state exceeds the value for superdeformation and is close to
the value 0.9 for hyperdeformation. 
In the $7/2^+_3$ state, we obtained the well-developed cluster 
structure like $^{9}$Li+$\alpha$(Fig.~\ref{fig:b13dense}(f)). 
In the final wave functions after the superposition, 
these two components of the largely deformed state (Fig.~\ref{fig:b13dense}(e)) 
and the $^{9}$Li+$\alpha$ cluster state (Fig.~\ref{fig:b13dense}(f))
constitute the rotational band, $K^\pi=1/2^+$, 
with a mixing of them.
The $|1/2^+_1\rangle$, $|3/2^+_3\rangle$, $|5/2^+_2\rangle$
states are dominated by $P^{J+}_{KM}\Phi_{\rm AMD}({\bf Z}^{1/2+}_1)$ 
in about 90\%, 65\% and 60\%, respectively. On the other hand,
the $|7/2^+_3\rangle$ and $|11/2^+_1\rangle$ contains major percentage of
$P^{J+}_{KM}\Phi_{\rm AMD}({\bf Z}^{7/2+}_3)\rangle$ with remarkable 
$^{9}$Li+$\alpha$ cluster structure, 
while other states in the $K^\pi=1/2^+$ band are the mixture of these
two components.
It indicates that the weak-coupling cluster feature is enhanced 
in high spin region of the $K^\pi=1/2^+$ band. 
The negative-parity states $P^{J-}_{KM}\Phi_{\rm AMD}({\bf Z}^{7/2+}_3)$
projected from the $^{9}$Li+$\alpha$ cluster structure
construct the $K^\pi=1/2^-$ band.

\subsection{Cluster feature in the $K^\pi=3/2^-$ and $K^\pi=1/2^+$}
In this subsection, we discuss cluster features of the deformed bands,
$K^\pi=3/2^-$ and $K^\pi=1/2^+$,
by analyzing the single-particle orbitals in the intrinsic states.
Hereafter, we mainly analze the major AMD wave functions, 
$\Phi_{\rm AMD}({\bf Z}^{5/2-}_1)$, $\Phi_{\rm AMD}({\bf Z}^{1/2+}_1)$ and 
$\Phi_{\rm AMD}({\bf Z}^{7/2+}_3)$ obtained by VAP for the 
$J^\pi_n=5/2^-_1(K^\pi=3/2^-)$, $J^\pi_n=1/2^+_1(K^\pi=1/2^+)$ and
$J^\pi_n=7/2^+_3(K^\pi=1/2^-)$ states, and compare them with that for the
intruder ground state $J^\pi_n=0^+_1$ of $^{12}$Be.

\begin{figure}[th]
\epsfxsize=8 cm
\centerline{\epsffile{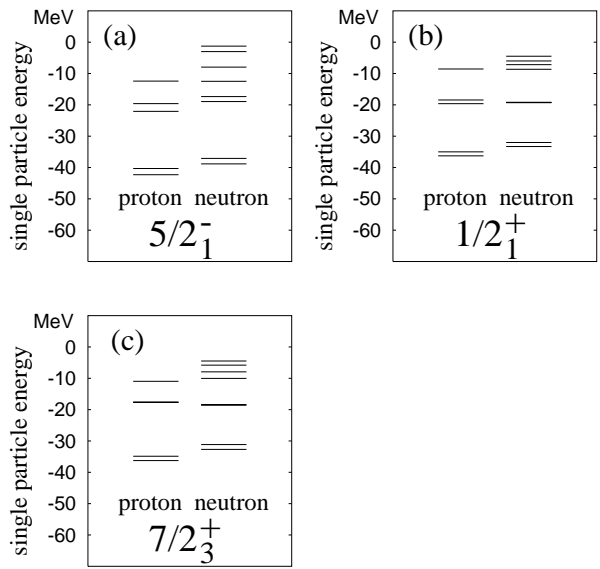}}
\vspace*{8pt}
\caption{
Single-particle energies in the $5/2^-_1$, $1/2^+_1$ and $7/2^+_3$
states.
The energies are calculated by diagonalizing
Hartree-Fock-like single-particle Hamiltonian within the single
AMD wave function \protect\cite{DOTE}
which dominates the $|J^\pi_n\rangle$ state.
\protect\label{fig:b13hfe}}
\end{figure}

First, we give single-particle energies in Fig.~\ref{fig:b13hfe}.
In these deformed states, the level structure of 
the single-particle wave functions 
shows a feature of the $2\alpha+p+4n$ structure rather than that of 
spherical shell structure.
That is to say, the lowest four proton orbitals and the lowest four 
neutron orbitals form the 2$\alpha$ core, while the higher orbitals correspond
to one valence proton and four valence neutrons.

\begin{figure}[th]
\epsfxsize=12 cm
\centerline{\epsffile{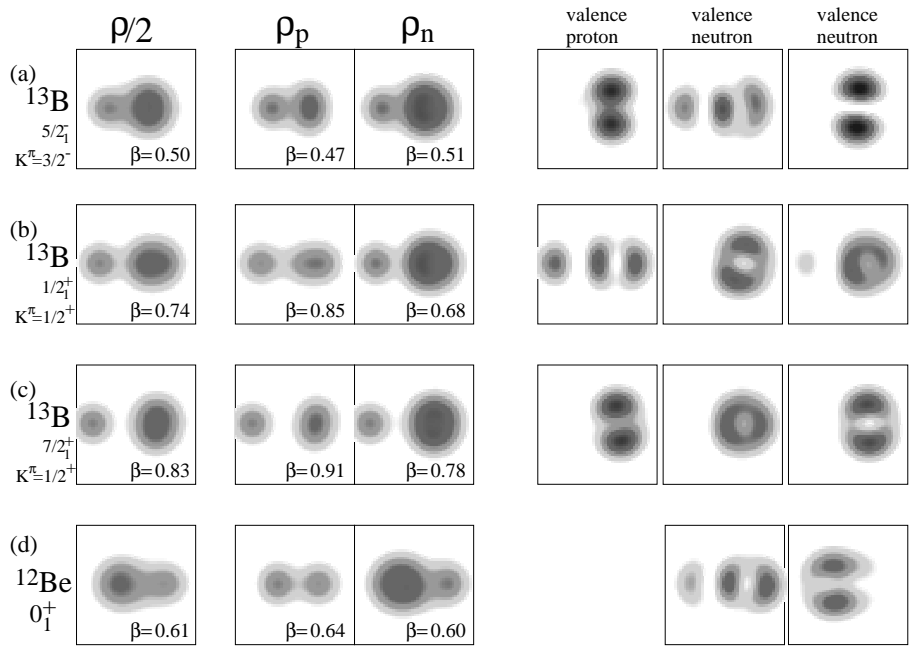}}
\vspace*{8pt}
\caption{Distribution of the proton, neutron and matter density, 
and density distribution of the valence nucleons.
The left three panels show the distribution and 
deformation parameter $\beta$ of matter, proton and neutron density.
In the right three panels, density distribution of the 
highest proton orbitals, the highest and the third neutron orbitals
are given. The single-particle orbitals are extracted by diagonalizing
Hartree-Fock-like single-particle Hamiltonian within a 
single AMD wave function 
\protect\cite{DOTE}.
The results are those calculated 
for the intrinsic AMD wave function $\Phi_{\rm AMD}({\bf Z}^{J\pi}_n)$, 
which dominates the $|J^\pi_n\rangle$ state.
The AMD results for the ground state of $^{12}$Be are also shown
for comparison.
The box size is 10 fm.
\protect\label{fig:b13single}}
\end{figure}

Next, we illustrate the density distribution of the single-particle 
wave functions for the valence nucleons in Fig.~\ref{fig:b13single}. 
The total densities of protons and neutrons are also given as well as
total matter densities in the figure.
Generally speaking, the matter densities show two-center structures in 
all these deformed states. 
However, the behavior of the valence nucleons are different among these
three states, $J^\pi_n=5/2^-_1(K^\pi=3/2^-)$, 
$J^\pi_n=1/2^+_1(K^\pi=1/2^+)$ and $J^\pi_n=7/2^+_3(K^\pi=1/2^-)$. 

In the $J^\pi=5/2^-_1(K^\pi=3/2^-)$ state (Fig.~\ref{fig:b13single}(a)), 
two valence neutrons occupy an
approximately positive-parity orbital, and the other two neutrons
and a proton are in orbitals with dominant negative-parity component.
Since the negative- and positive-parity orbitals of the valence nucleons
are associated with the $p$-orbitals and $sd$-orbitals, respectively, we can
roughly describe the states in the $K^\pi=3/2^-$ band 
by the neutron $2\hbar\omega$ excited configurations.
Let us turn to the molecular orbital features. 
The positive-parity orbital of the last two neutrons 
is largely deformed and has nodes along the 
$2\alpha$ direction(longitudinal axis).
This orbital well corresponds to the 
so-called $\sigma$ orbital in the molecular orbital picture
\cite{ITAGAKI,Oertzen-rev,Okabe77,SEYA,OERTZEN}.
It has been already known that the $\sigma$-like orbital 
of valence neutrons appear 
in various Be isotopes (see references in \cite{Oertzen-rev}).
In the AMD study\cite{Enyo-be12}, it has been revealed that 
the ground state of $^{12}$Be is dominated by 
the intruder state with two neutrons in the $\sigma$ orbital, which reduce
kinetic energy due to the developed $2\alpha$-core structure. 
In Fig.~\ref{fig:b13single}(4), density distribution and single-particle
orbitals of $^{12}$Be($0^+_1$) are shown.
As seen in the figure, the last two neutrons in the $^{12}$Be($0^+_1$)
state occupy the $\sigma$ orbital. The point is that the neutron structure
of the $^{13}$B($K^\pi=3/2^-$) band is very similar to that of the
$^{12}$Be($0^+_1$).
Therefore, we conclude that the $K^\pi=3/2^-$ is the band of the intruder 
neutron $2\hbar\omega$ states, and interpreted as $^{12}$Be($0^+_1$)+$p$, where
the $^{12}$Be($0^+_1$) has the intruder configuration and the 
additional proton strongly couples to the deformed core. 
It is also interesting that the additional proton 
in the normal $p$-shell affects a change of the deformation of total density, 
which results
in the smaller deformation of the $^{13}$B($K^\pi=3/2^-$) than 
the $^{12}$Be($K^\pi=0^+_1$).

In the band-head $1/2^+_1$ state of the $K^\pi=1/2^+$ band
(Fig.~\ref{fig:b13single}(b)), $1\hbar\omega$
excitation occurs in the proton shell. Namely, 
the last proton occupy a $\sigma$-like orbital, which 
is quite similar to that of the highest neutron orbital in the 
$K^\pi=3/2-$ band and also that in the $^{12}$Be($0^+_1$).
This is extraordinary configuration because excitations are naively
expected in the neutron side in case of neutron-rich nuclei.
It can be understood by the lowering mechanism of the 
$\sigma$ orbital due to the developed two-center structure.
In the two-center shell model\cite{twocenter}, the energy of the 
$\sigma$ orbital comes down with the increase of two-center distance $d$.
On the other hand, the negative-parity levels($\pi$)
originating in the $p$ orbitals split and
four of the negative-parity levels go up as the two-center distance increases.
As a result, the inversion of the positive-parity $\sigma$ orbital 
and the negative-parity $\pi$ orbitals occurs. Finally, in the large distance 
region $d\sim 5-6$ fm, 
the $\sigma$ orbital becomes the fifth orbital which is the
lowest one for valence nucleons around the $2\alpha$ core.
This situation is realized in the $1/2^+_1$ state, where 
the two-center structure with the distance $d\sim 5$ fm was obtained in the
present calculation.
It is also associated with the hyperdeformation, where the 
declined positive-parity orbital becomes the fifth lowest orbital
in the Nilsson's deformed shell model. In fact, the deformation $\beta=0.74$ 
of the $1/2^+_1$ state exceeds the value $\beta=0.6$ for the superdeformation.
On the other hand, four valence neutrons are localized 
in one side of the two-center structure. 
It is contrast to the molecular orbital feature of the valence proton, which 
is moving around the whole system.
It seems that 
spatial correlation of four valence neutrons is so strong that 
form a $p_{3/2}$ shell closure.
As a result, totally 8 neutrons separate into two groups 
consisting of 2 and 6 neutrons
with a weak-coupling feature.
We here give a comment on similarity of the neutron structure between 
the $1/2^+_1$ and $5/2^-_1$ states. 
Comparing the neutron structure of the $1/2^+_1$ with that of the
$5/2^-_1$, the profile of the total neutron density of the 
$1/2^+_1$ is similar to that of the $5/2^-_1$ described by 
the neutron $2\hbar\omega$ configuration. Therefore, we can propose 
an alternative interpretation in a mean-field picture that 
the $1/2^+_1$ state is roughly described by 
the $3\hbar\omega$ configuration with the proton $1\hbar\omega$ and
the neutron $2\hbar\omega$ excitations.
In the present calculation, we suggest 
that such the exotic state, the proton intruder
state of neutron-rich nuclei, may exist as the lowest $1/2^+$ state
at $E_x=8$ MeV.
Recently, the state at 4.83 MeV has been assigned to be a 
$1/2^+$ state by $^4$He($^{12}$Be,$^{13}$B$\gamma$)$X$ 
experiments\cite{Ota07}. Because of strong production 
via the proton-transfer to the $^{12}$Be($0^+$) state and analysis of 
angular dependence, Ota {\it et al.}
suggested this $1/2^+$ state to be a proton intruder state. 
The present prediction of the proton intruder configuration 
in the $1/2^+_1$ state is consistent with this observation
though the theoretical excitation energy of the present calculation 
is slightly higher than the experimental value.

In the $7/2^+_3$ state
 in the $K^\pi=1/2^+$  band, all the valence proton and
the valence four neutrons are localized around one of the $\alpha$ 
cores(Fig.~\ref{fig:b13single}(c)).
It indicates that the molecular orbital aspect disappears, 
while the weak-coupling feature of $\alpha$-$^9$Li clustering 
is enhanced. Because of the spatial localization, the orbitals of the
valence nucleons have no definite parity. 
In this case, the unnatural parity '$+$' of the total system
is carried by the parity asymmetric $\alpha$-$^9$Li structure.
It is in different situation from the $1/2^+_1$ state with the intruder proton
configuration, where the parity '$+$' originates in the proton $1\hbar\omega$ 
excitation of the single-particle orbital from negative to positive one.
As mentioned before, after the superposition of the AMD wave functions,
the proton intruder state and the $\alpha+^9$Li cluster state are mixed
to each other to contribute the structure change depending on $J$
in the $K^\pi=1/2+$ band.
The former component is dominant 
in the low spin states, while the latter component
is significant in high spin states.

\section{Summary}\label{sec:summary}
The excited states of $^{13}$B were studied with a method of 
antisymmetrized molecular dynamics(AMD). 
We obtained the largely deformed states
which construct the rotational bands, $K^\pi=3/2^-$, $K^\pi=1/2^+$ and $K^\pi=1/2^-$. 
In these deformed states, we found various kinds of cluster aspect
including molecular orbital features in the two-center structure.

The $K^\pi=3/2^-$  band is the deformed one
with molecular orbital structure.
This band is described by the intruder 
neutron $2\hbar\omega$ configuration, and interpreted as $^{12}$Be($0^+_1$)+$p$, where
the $^{12}$Be($0^+_1$) has the intruder configuration and the 
additional proton strongly couples to the deformed $^{12}$Be core. 
The excited neutron orbital is regarded as the $\sigma$ orbital 
in the molecular orbital picture.
Experimentally, there is a report which has been suggested the excited state
at 10 MeV to be a high spin state of the neutron $2\hbar\omega$ configuration
\cite{Kalpakchieva00},
which would be a member of this $K^\pi=3/2^-$ band.

We found the proton intruder structure
in the $K^\pi=1/2^+$ band with remarkably large deformation.
The deformation $\beta=0.74$ of the band-head $1/2^+_1$ state
is larger than  
that for the superdeformation. In the molecular orbital picture, 
the last proton is described by the $\sigma$ orbital. 
This is very exotic state with the proton intruder configuration
in neutron-rich nuclei. In the present calculations, it was suggested
that the $K^\pi=1/2^+$ band starts from the
lowest $1/2^+$ state at $E_x=8$ MeV. 
We assigned this state to the recently observed 
$1/2^+$ state at 4.83 MeV, which has been suggested to be  
the proton intruder state by Ota {\it et al.} 
in the $^4$He($^{12}$Be,$^{13}$B$\gamma$)$X$ experiments.
In the high spin states of the $K^\pi=1/2^+$ band, $\alpha$+$^9$Li-like
cluster structure well develops.

It is striking that such the deformed states with 
highly excited configurations or 
well-developed cluster states appear in the energy region compatible to
the normal excited states, even though
$^{13}$B has the neutron-shell closure in the ground state. 
Especially the 
molecular orbital, $\sigma$, is formed in the deformed states of $^{13}$B
as well as neutron-rich Be isotopes.
The cluster aspect of the deformed states in $^{13}$B can be 
understood in natural extension of cluster structure of Be isotopes.
That is to say, the formation of the $2\alpha$-cluster core and the role 
of the valence nucleons(one proton and four neutrons) 
are key in the largely deformed states of $^{13}$B.
It is challenging to 
investigate possible exotic structure with clustering
in excited states of further 
neutron-rich B isotopes like $^{15}$B and $^{17}$B.

\section*{Acknowledgments}

The authors would like to thank Dr. Ota and Prof. Shimoura 
for the valuable discussions. In fact, 
this study has been motivated by their 
suggestion of the ``proton intruder state''.
They are also thankful to members of 
Yukawa Institute for Theoretical Physics(YITP)
and Department of Physics in Kyoto University.
The computational calculations in this work were performed by the 
Supercomputer Projects 
of High Energy Accelerator Research Organization(KEK)
and also the super computers of YITP.
This work was supported by 
Grant-in-Aid for Scientific Research 
Japan Society for the Promotion of 
Science and a Grant-in-Aid for Scientific Research from JSPS.
It is also supported by the Grant-in-Aid for 
the 21st Century COE "Center for Diversity and Universality in Physics"
from MEXT.
Discussions during the workshop YITP-W-06-17 on Nuclear Cluster Physics
held in YITP were useful to complete this work.

\end{document}